\documentclass[10pt,conference]{IEEEtran}
\IEEEoverridecommandlockouts
\usepackage{cite}
\usepackage{amsmath,amssymb,amsfonts}
\usepackage{algorithmic}
\usepackage{graphicx}
\usepackage{textcomp}
\usepackage{xcolor}
\def\BibTeX{{\rm B\kern-.05em{\sc i\kern-.025em b}\kern-.08em
    T\kern-.1667em\lower.7ex\hbox{E}\kern-.125emX}}

\usepackage{array}
\usepackage{wrapfig}
\usepackage{multirow}
\usepackage{tabularx}
\usepackage{tcolorbox}
\usepackage{balance}
\usepackage[numbers]{natbib}

\usepackage{blindtext}
\usepackage{hyperref}

\begin{document}

\title{APIContext2Com: Code Comment Generation by Incorporating Pre-Defined API Documentation*\thanks{This research is support by a grant from the Natural Sciences and Engineering Research Council of Canada RGPIN-2019-05175.}}

\author{\IEEEauthorblockN{Ramin Shahbazi}
\IEEEauthorblockA{\textit{Department of Computer Science} \\
\textit{University of British Columbia}\\
Okanagan, Canada \\
ramin20@mail.ubc.ca}

\and

\IEEEauthorblockN{Fatemeh Fard}
\IEEEauthorblockA{\textit{Department of Computer Science} \\
\textit{University of British Columbia}\\
Okanagan, Canada \\
fatemeh.fard@ubc.ca}

}

\maketitle
\thispagestyle{plain}
\pagestyle{plain}

\begin{abstract}

Code comments are significantly helpful in comprehending software programs and also aid developers to save a great deal of time in software maintenance. Code comment generation aims to automatically predict comments in natural language given a code snippet. 
Several works investigate the effect of integrating external knowledge on the quality of generated comments. In this study, we propose a solution, namely APIContext2Com, to improve the effectiveness of generated comments by incorporating the pre-defined Application Programming Interface (API) context. The API context includes the definition and description of the pre-defined APIs that are used within the code snippets. As the detailed API information expresses the functionality of a code snippet, it can be helpful in better generating the code summary. We introduce a seq-2-seq encoder-decoder neural network model with different sets of multiple encoders to effectively transform distinct inputs into target comments. A ranking mechanism is also developed to exclude non-informative APIs, so that we can filter out unrelated APIs. We evaluate our approach using the Java dataset from CodeSearchNet. The findings reveal that the proposed model improves the best baseline by 1.88 (8.24\%), 2.16 (17.58\%), 1.38 (18.3\%), 0.73 (14.17\%), 1.58 (14.98 \%) and 1.9 (6.92 \%) for BLEU1, BLEU2, BLEU3, BLEU4, METEOR, ROUGE-L respectively. Human evaluation and ablation studies confirm the quality of the generated comments and the effect of architecture and ranking APIs.

\end{abstract}

\section{Introduction}

Code comment generation aims to generate readable summaries, which describe the functionality of the source code. Comments are crucial in comprehending the source code with the least possible time and effort \cite{ko2006exploratory,latoza2006maintaining}. Software developers spend a lot of time understanding code, especially during maintenance.  Much time could be saved with proper documentation and summaries available \cite{sridhara2010towards,xia2017measuring}. On the other hand, manually providing proper comments is time-consuming and sometimes tedious. Additionally, comments need to be updated as software projects evolve, otherwise, they could become misleading as they may fail to properly describe the true functionality of the code snippet over time. In many cases, comments are missed or outdated with software growth \cite{moreno2013automatic,singer2010examination}. Thus, it is crucial to leverage automatic techniques to generate comments for code snippets. 

Several studies have adopted rule-based approaches such as template generating \cite{sridhara2010towards} and Information retrieval (IR) \cite{eddy2013evaluating,haiduc2010supporting,haiduc2010use, rodeghero2014improving, wong2015clocom} techniques for code comment generation. The former approach provides some pre-defined templates to help developers produce readable comments. Despite that, the templates are project-specific and developers need to spend time understanding and adapting these templates. Moreover, producing the templates for different projects requires extensive domain knowledge. The IR approaches generate comments by selecting tokens from the source code, or by re-using comments of similar source codes. However, there are two limitations to the IR approach. First, it fails to provide meaningful comments in the case of poor naming such as irrelevant method names. Second, similar code snippet may not exist within the repository.

Researchers have adopted neural machine translation (NMT) for code comment generation, and the results are promising \cite{iyer2016summarizing, hu2018deep, wan2018improving, movshovitz2013natural, bahdanau2014neural, haije2016automatic}. The neural network models usually consist of an encoder-decoder framework, following a variant of RNN architecture. Iyer et al. \cite{iyer2016summarizing} proposed a seq2seq Long Short Term Memory (LSTM) model where the encoder receives source code as input and the decoder generates comment out of the encoder's output using an attention mechanism. Many studies \cite{hu2020deep, leclair2019neural, wan2018improving} leveraged Abstract Syntax Tree (AST) to better incorporate syntactical information of source code.
 The findings of various studies confirm the strength of neural frameworks on code comment generation tasks. However, code comment generation can still be improved by incorporating some hidden external information. 
 Recently, several works studied the effect of integrating external information into model to boost the performance of the model. Retrieving  similar source code and the corresponding comments, and leveraging the comments as an external input to improve the model performance are such examples \cite{wei2019retrieve, zhang2020retrieval}. 

Application Programming Interface (API) is also used in previous works as an external input to improve the generated comments \cite{hu2018summarizing, shahbazi2021api2com}.
Source code functionality can be defined by a sequence of pre-defined APIs used inside the code  \cite{hu2018summarizing}. Developers often apply a series of related APIs in their code for a specific task \cite{hu2018summarizing}. Although source code might contain different style conventions, API sequences are often similar for a particular feature. For instance, a regular way to write a piece of code to produce content in a JSON file would be applying the following sequence of APIs, \verb|FileWriter.write| and \verb|FileWriter.flush|. 
Each pre-defined API owns a full definition and description, which can be found in the JDK reference documentation. In this example, the descriptions for the APIs mentioned above are\footnote{https://docs.oracle.com/javase/7/docs/api/java/io/OutputStreamWriter.html}:  \textit{``Writes a portion of a string”} and \textit{``flushes the stream”}, respectively. We can see that the descriptions are defining the functionality of the main method. 
Therefore, we suppose that extracting the APIs documentation and incorporating them into the code comment generation task can be helpful and improve the performance of the model. 

Hu et al. \cite{hu2018summarizing} leverages API sequence summarization task to transfer knowledge from API name sequences to code summarization task. Shahbazi et al. \cite{shahbazi2021api2com} incorporates API description to improve the generated comments. However, they fail to properly capture the knowledge from API documentation. They conclude that as the number of APIs in a method increases, the performance of the model in predicting the comments decreases. They hypothesized that when the number of APIs increases,
it adds noise to the model and misleads the model to generate better comments. We believe this is due to simply concatenating API descriptions text in advance and feeding the model with long sequential text.

In this paper, we propose \textbf{APIContext2Com}, where we leverage new ideas to cover the problems of the previous work \cite{shahbazi2021api2com}. 
First, in addition to the API description, we also incorporate API \textit{definition} which includes  the \textit{API parameters' types and names}. API definition enriches the model by grasping some hidden information about APIs which are helpful in generating comments. 
Second, we propose a novel architecture which lets the model process each API description and definition in different encoders.
Third, we develop a ranking mechanism to identify informative APIs, thus removing APIs which are not useful for generating the code comments. This ranking mechanism is proposed as some APIs, such as \verb|toString()|, are usually general/unrelated to the method and they do not add any information to the model. 
On the other hand, some APIs are far more specific and provide detailed knowledge about the main functionality of the given code snippet, thus are helpful in generating comments. The ranking mechanism is especially useful in cases where a method contains more than three APIs. 

APIContext2Com leverages four different inputs for each method: source code, AST, a set API descriptions, and a set of API definitions. The architecture contains two single encoders and two sets of encoders where the first two encoders receive source code and AST, and two sets of encoders receive API descriptions and definitions.  
We perform our study on a large-scale Java corpus containing 137,007 samples collected by Husain et al. \cite{husain2019codesearchnet}. We further conduct several experiments to indicate the effect of removing unrelated APIs using the ranking mechanism as well as adding API documentation to the model. Our finding shows that removing low rank APIs enhances the quality of the generated code.
APIContext2Com improves the results of the best baseline model by 1.88 (8.24\%), 2.16 (17.58\%), 1.38 (18.3\%), 0.73 (14.17\%), 1.58 (14.98 \%) and 1.9 (6.92 \%) scores for BLEU1, BLEU2, BLEU3, BLEU4, METEOR, ROUGE-L, respectively.

The contributions of our work are as follow: 

\begin{itemize}
  \item We incorporated the API definition including the API parameters' names and their types.
  \item We develop a ranking mechanism to score the APIs based on their similarity to the main method, and only include the relevant APIs to the model.
  \item We leverage a different architecture to capture the knowledge from APIs context. 
\end{itemize}




The rest of this paper is organized as follows:

In the next section, we present two examples explaining the motivation of this study. Next, we elaborate on the details of the new approach in section \ref{sec:approach} followed by the experiments and results in Sections \ref{sec:experiments} and \ref{sec:results} respectively. 
Section \ref{sec:discussions} discusses different aspects of the experiment results. Next we mention the threats to validity in Section \ref{sec:threats} and review the related studies in Section \ref{sec:relatedWork}. Finally, we conclude the paper in Section \ref{sec:conclusion}.

\begin{figure}[htbp]
\centerline{\includegraphics[width=1\linewidth]{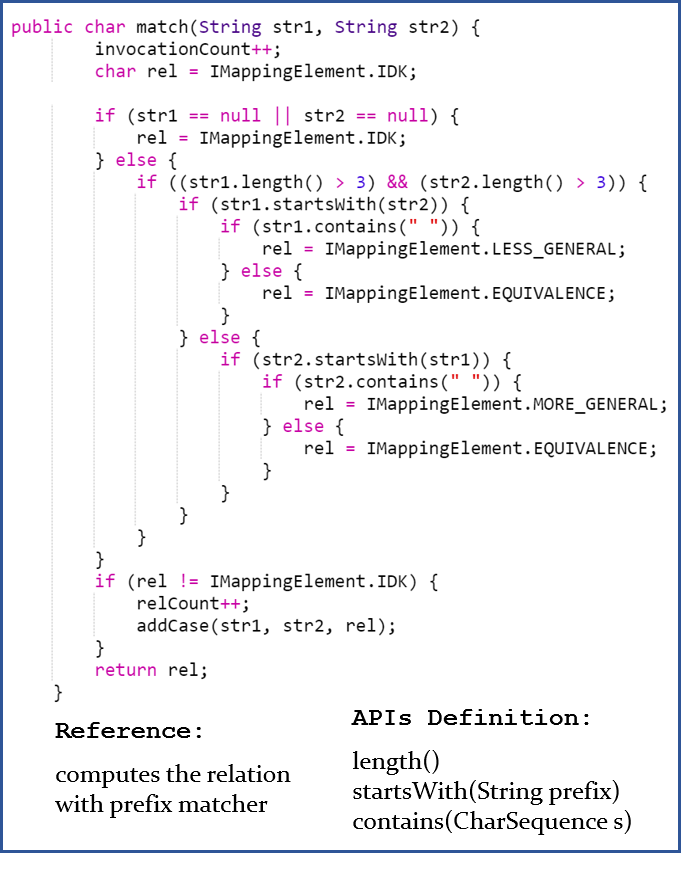}}
\caption{ API definition, i.e., its parameters' types and names, can help generating code comments. The word `prefix' is used in the reference summary and in the definition of the second API used in the method, but it is not mentioned in the code snippet.}
\label{fig:motivation-APIDefinition}
\end{figure}

\begin{figure*}
  \includegraphics[width=\textwidth]{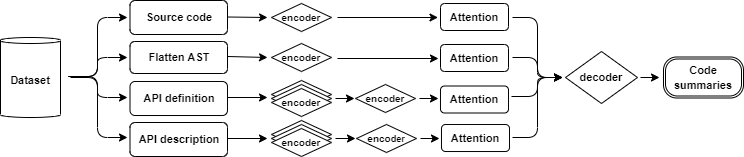}
  \caption{ An overview of APIContext2Com architecture}
   \label{fig:architecture}
\end{figure*}

\section{Motivation example}
In this section, we present two examples from the dataset where the first one explains how API definition is helpful to the model and the second one elaborates on the ranking mechanism and the strategy which is applied to exclude non-related APIs. 

API definition consists of API name followed by parameters' types and names. Inside a code snippet, the information about API parameters is missing. APIs are only called, and desired data are passed to them. However, looking up the APIs in JDK reference documentation can provide the exact definition of parameters and, more importantly, meaningful names which can suggest useful knowledge. 
Fig. \ref{fig:motivation-APIDefinition} illustrates a Java method employing three internal APIs, where the source code, its reference summary, and the definition of the APIs it calls are presented. 
The reference summary is ``computes the relation with prefix matcher" where the word ``prefix" is used. Exploring the code snippet reveals that this word is not mentioned within the code.
Interestingly, this word can be found as the defined parameter of the second applied API, \verb|startsWith(string prefix)|. It is a solid clue signifying that integrating API definitions are helpful and so we hypothesize that the API definitions can increase the model's performance.

\begin{figure}[htbp]
\centerline{\includegraphics[width=1\linewidth]{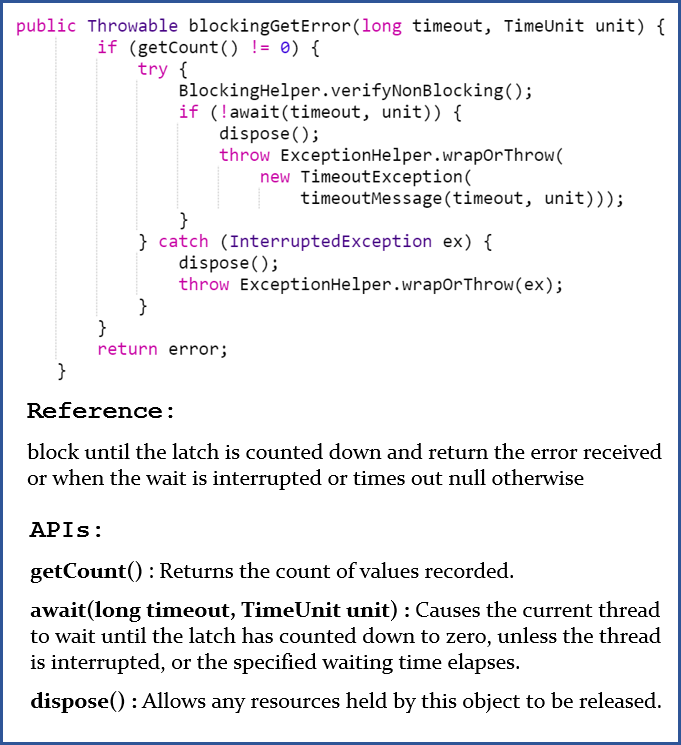}}
\caption{ Motivation example for ranking the APIs. The parameters of the second API are similar to the ones of the code snippet. But the other two APIs do not have similar parameters with the code sample. Thus the definition of the second API can be useful to generate the code summary, but this is not true for the other two APIs.  }
\label{fig:motivation-similarity}
\end{figure}

Fig. \ref{fig:motivation-similarity} depicts a Java method followed by reference summary and the definitions and descriptions of the APIs. A quick look at the APIs reveals that the second API owns a description that is highly similar to the reference summary. Similarly, they have the exact same input parameters. This observation motivates that containing similar input parameters can indicate possible similarity in functionality and consequently similar summaries. 
However, not all the APIs in a method are helpful to be included in the model. Many APIs such as \verb|tostring()| are irrelevant to the main functionality of the given code snippet. In these cases, neither the description nor the definition provides practical knowledge to the model. 
Moreover, Shahbazi et al. \cite{shahbazi2021api2com} concluded that API documentation is not helpful for the methods with more than three APIs, and it misled the model in providing the summaries. 
Therefore, we develop a ranking mechanism to calculate the similarity between each API used in the method and the given method by computing a score for each API. To achieve this goal, we compare the input parameters of the API and the method, and with each difference, one negative point is given to the API. In this way, we can identify the APIs which are more likely to provide useful information for the code summarization. In the example of Fig. \ref{fig:motivation-similarity}, the \verb|await(long timeout,TimeUnit unit)| is ranked the highest, but the other two APIs, \verb|getCount()|, \verb|dispose()|, receive two negative points as they are differences in their parameters and depending on the number of APIs we are including they have lower chance to be used in generating comments. 


\section{Proposed approach} \label{sec:approach}

\subsection{Model Overview} 

Following \cite{leclair2019neural}, we propose a Sequence-2-Sequence architecture with multiple encoders for this study. Shahbazi et al. \cite{shahbazi2021api2com} propose an architecture with three different encoders each one receiving one input including source code, AST and API description. For API description, they concatenate all the different descriptions and use it as one input to a single encoder. As each description explains the functionality of a distinct API, concatenation of them before passing it to the model could create a long and meaningless sequence of tokens. 
Thus, in this study, we are applying a list of encoders for the API descriptions and API definitions, each API description or definition is processed by one encoder and the final hidden state of all encoders are used as input to a final encoder. Fig. \ref{fig:architecture} depicts the overall architecture of our proposed model. 

Code is fed to a single encoder as shown in Fig. \ref{fig:architecture}. The encoder contains an embedding layer and a GRU layer. We apply encoder is applied for AST. First, the AST tree representation of the source code is generated using srcMl tool, and then it is flattened using SBT method \cite{hu2018deep}. The flattened AST is used as a sequence of plain text to feed the encoder. 
API description/API definition are different inputs than code/AST. We only have one code and one AST for each method while a list of APIs exists in each method. Thus, each API is fed to one encoder and then the hidden state of encoders are used to feed the final encoder. The output of the final encoder is the same as the code and AST encoders' output. As we already mentioned, we define a threshold number, so we only integrate a specific number of APIs based on the ranking mechanism. 
Next, the attention for each encoder is calculated similar to a typical encoder-decoder models. For AST and code, the attention weights are computed for each token, while for APIs the attention score is calculated per API. Next the encoders' hidden states are concatenated and passed to a fully-connected layer and then to the decoder.

\subsection{Model Details}

\textbf{Encoder:} Each encoder receives one type of text sequence from  different inputs, including a code snippet denoted by $code$, flattened AST denoted by $ast$, a sequence of API definition denoted by $def$ and a sequence of API description denoted by $des$. The architecture of encoders for code and AST are similar. We consider the code snippet encoder to elaborate on the structure below. Initially, the encoder maps the token sequence to a  token embedding. Then, the embedding output is passed to a bidirectional GRU to extract the contextual information. GRU framework is designed and optimized to prevent the process from long-term dependency issues \cite{DBLP:journals/corr/ChungGCB14}. At each step, the hidden state of the GRU is defined as follows:

\begin{equation}
\begin{aligned}
h_t^{code} = GRU(x_t^{code}, h_{t-1}^{code})
\end{aligned}
\end{equation}

Where $x_t^{code}$ is code input to the model (code sequence) at step t, and $h_{t-1}^{code}$ denotes the hidden state for code input at step $t-1$.

The hidden state for AST is obtained in a similar way as follow:

\begin{equation}
\begin{aligned}
h_t^{ast} = GRU(x_t^{ast}, h_{t-1}^{ast})
\end{aligned}
\end{equation}

Where $x_t^{ast}$ and $h_t^{ast}$ are flattened AST input and hidden state respectively. 

As each method can have multiple APIs, which means multiple API definitions and descriptions, instead of concatenating the descriptions as done in \cite{shahbazi2021api2com}, we map each API to a single encoder. We consider API definition to elaborate on the process. A single encoder processes each single API definition, resulting in a unique hidden state for each API as follow:

\begin{equation}
\begin{aligned}
h_t^{def_1} = GRU(x_t^{def_1}, h_{t-1}^{def_1}) \\
h_t^{def_2} = GRU(x_t^{def_2}, h_{t-1}^{def_2}) \\
\vdots\\
h_t^{def_n} = GRU(x_t^{def_n}, h_{t-1}^{def_n}) 
\end{aligned}
\end{equation}

Further, the hidden states are used as a sequence of tokens to a final encoder, resulting in a single hidden state representing the entire API definition.
The final encoder does not include an embedding layer as the inputs are already numeric vectors. It is formulated as below:

\begin{equation}
\begin{aligned}
h_t^{def} = GRU(h^{def_t}, h_{t-1}^{def})
\end{aligned}
\end{equation}

Where $h_t^{def}$ is the final hidden state of API definition. Similarly, we can define all the formulas for API description ($h_t^{des}$).

\textbf{Attention:} Attention is a component that enables the decoder to attend to relevant tokens more effectively \cite{sutskever2014sequence}. The attention mechanism helps the decoder focus on proper tokens within different inputs at each prediction step. For instance, the keyword \verb|return| or \verb|set| in code can represent words such as ``get" or ``update" in the comment, respectively. As a result, different parts in comments can be related to distinct parts of the code. 
Let's consider the relation between the final comment and the source code. In order to compute the attention weights for target token $y_i$, the decoder hidden state of  the previous token,  $h_{i-1}^{'}$,  is used as follow,

\begin{equation}
\begin{aligned}
\alpha_{ij} = a(h_{i-1}^{'},h_j^{code})
\end{aligned}
\end{equation}

Where $a$ is alignment model, $h_{i-1}^{'}$ is decoder hidden state and $\alpha_{ij}$ denotes how related the code input at position $j$ and the target comment at position $i$ are. The context vector, further, is defined as a weighted sum of hidden states of inputs,

\begin{equation}
\begin{aligned}
c_i^{code} = \sum_j(\alpha_{ij} h_j^{code}) 
\end{aligned}
\end{equation}

Similarly, the context vector is calculated for the other three inputs. 

\textbf{Decoder:} The decoder component aims to generate final comments in human language. The following formula is used to generate the output and the decoder hidden state at time step t:
 
\begin{equation}
\begin{aligned}
h_t^{'} = GRU(h_{t-1}^{'}, y_{t-1} )
\end{aligned}
\end{equation}

$h_{t-1}^{'}$ denotes the decoder hidden state for the previous token, and $y_{t-1}$ is the previous token. The initial hidden state to the decoder is a combination of the final encoder hidden state for all four inputs. To calculate the initial hidden state, the four encoders' hidden states are concatenated and passed on to a fully connected layer: 

\begin{equation}
\begin{aligned}
h_0^{'} = W_c[h^{code};h^{ast};h^{des};h^{def}]+B_c 
\end{aligned}
\end{equation}

$W_c$ and $B_c$ represent trainable parameters. The final context vector is also calculated in a similar way. 
\begin{equation}
\begin{aligned}
c_t = W_c[c^{code};c^{ast};c^{des};c^{def}]+B_c 
\end{aligned}
\end{equation}
Next, the probability of the target token at time t is computed based on the previous token, hidden state, and context vector. 
\begin{equation}
\begin{aligned}
p(y_t|y_{t-1}, ..., y_1,x)= g(y_{t-1},h_t^{'}, c_t)
\end{aligned}
\end{equation}

Here, $g$ denotes the softmax activation function.

We study the effect of using another Recurrent Neural Network (RNN)-based architecture, namely, Long Short Term Memory (LSTM) instead of GRU as well in Section \ref{sec:gru-effect}.

\subsection{Ranking System}
As mentioned earlier, the list of APIs that are fed to the model are based on a ranking system; meaning that we can rank the APIs used in the function and feed the $n$ top-ranked ones to the model. The ranking is based on the number of differences among the parameters of the function and the APIs. Let $P=\{p_1,...p_n\}$ be the list of function's parameters and $M_i=\{m_i1,...m_ik\}$ be the list of parameters of $API_i$ (API\textsubscript{i} is used in the function). The score $S_i$ for $API_i$ is defined as 

\begin{equation}
\begin{aligned}
S_i = - Diff(P,M_i)
\end{aligned}
\end{equation}

where $Diff$ calculates the number of differences among $P$ and $M_i$. This score is calculated for all the APIs used in the function. The API with higher $S_i$ value is the highest rank.

\section{Experiments} \label{sec:experiments}
\subsection{Dataset}

In this study, we employ CodeSearchNet \cite{husain2019codesearchnet} dataset to perform the experiments, including training and testing the proposed model as well as running the baselines. Many recent studies have leveraged this well-known dataset to conduct experiments, especially in code comment generation \cite{feng-etal-2020-codebert,wang2020cocogum}. CodeSerachNet consists of a set of different dataset sources from Github repositories. These datasets are collected using a tool named Libraries.io which aims to identify the projects with higher validity. All the projects without a valid license have been removed from the final dataset. This dataset contains six different languages, including Python, Javascript, Ruby, Go, Java, and PHP. For the purpose of our study, we extracted the Java portion of the dataset to run the experiments. Each data row includes a complete Java method and corresponding reference summary, which are required for this study. 
Similar to \cite{leclair2019neural}, we perform some pre-processing steps before training the model. Tokens are split where camelcase or snake case exists. Further, we remove the punctuation from the code and convert all tokens to lower case.
Similar to \cite{feng-etal-2020-codebert} study, we set the maximum length of code and summary to 256 and 64, respectively. We removed the records with long text where AST cannot be generated using truncated code, and we might end up losing valuable knowledge. This pre-processing step is required as a few models in this study apply AST, and we have to provide an equal condition to have a fair comparison. The dataset is split into train, validation, and test, as shown in Table \ref{table:Datasetstatistics}. Train and validation sets are utilized for training the model and selecting the best model, while the test set is used to evaluate the model.

\begin{table}[h!]
\caption{Dataset statistics} 
\label{table:Datasetstatistics}
\small	
\begin{tabular}{ p{0.2\linewidth}p{0.3\linewidth}p{0.3\linewidth} }
 \hline
Dataset & Records before filtering  & Records after filtering\\
\hline
Train & 164,923 & 137,007 \\
Validation &	5,183 & 4,326    \\
Test &	10,955 &	9,011\\

 \hline
\end{tabular}
\end{table}

\subsection{API Documentation Extraction}
We modified the scraper developed by \cite{shahbazi2021api2com}, to download the API definition from Java Documentation Kit (JDK) . API definition contains all the parameters' types and names. In order to match the APIs within the methods and downloaded APIs context, we first extract all the APIs from the given methods. We used SrcML tool \footnote{https://www.srcml.org/} to convert the source code to AST representation. Then we used it to identify the APIs through AST representation. After matching the identified APIs with scraped APIs, we extracted the names and types of the parameters.

\subsection{Experiment Setting}
We implement the proposed approach on PyTorch framework\footnote{https://pytorch.org/}. For training the model, the embedding and hidden state sizes are set to 256 and 512, respectively. We set the batch size and number of iterations (epochs) to 32 and 100. However, the training will stop in case no improvement is observed in seven consequent iterations. \textit{Stochastic gradient descent} is applied as the optimizer with the config of 0.1 for the initial learning rate. In order to reduce the possibility of overfitting, multiple dropout layers with the value of 0.2 are adopted. The model is trained on the training set, the best model weights over different iterations are identified through the validation set, and finally, the selected model is evaluated on the test set.

\begin{table*}[h!]
\caption{BLEU, METEOR, and ROUGE-L scores of APIContext2Com and other models. The highest scores are bold.}
\label{table:RQ1}
\small	
\begin{tabularx}{\textwidth}{ p{0.21\textwidth}p{0.1\textwidth}p{0.1\textwidth}p{0.1\textwidth}p{0.1\textwidth}p{0.1\textwidth}p{0.12\textwidth}  }
 \hline
 Models &\multicolumn{6}{c}{Metrics} \\
 \cline{2-7}
 & BLEU1 (\%) & BLEU2 (\%) & BLEU3 (\%) & BLEU4 (\%) & METEOR(\%) &  ROUGE-L (\%) \\
 \hline
 
 DeepCom & 17.25 &8.90 & 4.83 & 2.72 &  9.5 & 21.06 \\

 AST-Attendgru   &22.02  &9.97  &5.21  &3.20  & 9.45 & 23.06\\
 TL-CodeSum&   21.65  & 10.10   &5.69&3.74&8.36&22.10\\
 Rencos &22.82  & 12.29&  7.54 &5.15&10.55&27.46\\

 APIContext2Com&   \textbf{24.70}  & \textbf{14.45} & \textbf{8.92} & \textbf{5.88} & \textbf{12.13} & \textbf{29.36}  \\

 \hline
\end{tabularx}

 \end{table*}

\section{Experiment Results} \label{sec:results}

\subsection{Baselines} 

In this section, we explore the baselines details and their architecture. We consider four state-of-the-art baselines as follows:

\textbf{Rencos} \cite{zhang2020retrieval} The authors of Rencos propose an attentional encoder-decoder model with the assistance of retrieval-based summarization task. They improve the model by retrieving similar methods and incorporating them to the model. Initially, the model is trained on the training set consisting of code and summary pairs. Further, during the test phase, two most similar methods are retrieved from the train set in terms of semantic and syntactic. The code and the retrieved methods are encoded, and the decoder generates the final summary by fusing the inputs. They include Bi-LSTM layers as the variant of RNN in their model.

\textbf{TL-CodeSum} \cite{hu2018summarizing} introduces a sequence-2-sequence model with GRU framework, which adapts learned API knowledge from a different but related task. In fact, they develop a code comment generation model with the assistance of the knowledge transferred from an API sequence summarization model. The latter model intends to establish a mapping between API sequence and the respective summaries.  The knowledge learned in API sequence summarization task is transferred to the code comment generation model to enhance the performance of the model.

\textbf{AST-AttendGRU} \cite{leclair2019neural}
proposes a GRU encoder-decoder code comment generation model with an attention mechanism. They incorporate flattened AST sequence to the model to effectively capture the syntactical information of the code snippet. They leverage two different encoders where one processes code snippet tokens and the other one AST tokens. In this way, both semantical and syntactical information of the code snippet can be taken into account to improve the generated comments.  

\textbf{DeepCom} \cite{hu2020deep}
DeepCom introduces a new approach to combine abstract syntax tree into the model. They design a tree traversal technique, namely Structure-Based (SBT) Traversal, which is able to capture the syntactical information of the source code effectively. Further, they flatten the AST tree representation using SBT and incorporate AST into the model in the form of a text sequence. 

We run all these models according to the best hyperparameters reported by their authors. Note that pre-trained models such as CodeBERT \cite{feng-etal-2020-codebert} and GraphCodeBERT \cite{DBLP:conf/iclr/GuoRLFT0ZDSFTDC21} are not chosen here for results' comparison, as i) they have a different paradigm which is language modeling through pre-training on large corpus of data, ii) they are based on the Transformer \cite{vaswani2017attention} architecture, and iii) they are larger models and it is shown that the larger the language models, the better their performance is \cite{devlin-etal-2019-bert}. Additionally, similar works do not compare them with their approach \cite{wei2019retrieve, zhang2020retrieval, li2021editsum}. As this is not a fair comparison to include such language models, we only compare APIContext2Com with RNN-Based \cite{zhang2020retrieval} approaches which are not language models.

\begin{table*}[h!]
\caption{effect of ranking mechanism and number of APIs on model performance. The highest scores are bold.}
\label{table:RQ2}
\small	
\begin{tabularx}{\textwidth}{ p{0.28\textwidth}p{0.09\textwidth}p{0.09\textwidth}p{0.09\textwidth}p{0.09\textwidth}p{0.09\textwidth}p{0.11\textwidth}  }
 \hline
 Models &\multicolumn{6}{c}{Metrics} \\
 \cline{2-7}
 & BLEU1 (\%) & BLEU2 (\%) & BLEU3 (\%) & BLEU4 (\%) & METEOR(\%) &  ROUGE-L (\%) \\
 \hline

2 APIs  
&24.02&	13.98&	8.65&	5.65&	11.64&	28.33\\

3 APIs&   \textbf{24.70}  & \textbf{14.45} & \textbf{8.92} & \textbf{5.88} & \textbf{12.13} & \textbf{29.36}  \\

4 APIs&   24.43  & 14.24 & 8.85 & 5.84 & 11.95 & 28.98  \\
 
 All the APIs&   23.91  & 13.92&8.58&5.73&11.84&28.52\\

 \hline
\end{tabularx}

 \end{table*}

\subsection{Evaluation Metrics} 

Following previous similar studies \cite{ahmad2020transformer, wan2018improving, zhang2020retrieval} we employ three well-know metric including BLEU 1-4  \cite{papineni2002bleu}, ROUGE-L \cite{lin2004rouge} and METEOR \cite{banerjee2005meteor} to evaluate our proposed approach. These metrics are widely applied in code comment generation tasks to evaluate the similarity of the generated comment and reference comment. In other words, the specified metrics signify the ability of the models to predict similar summaries. 

BLEU measures the precision of the average n-grams between the reference and generated summaries with the help of a brevity penalty  \cite{papineni2002bleu}, where n shows the n-grams and is reported for $n \in [1,4]$. 
The brevity penalty penalizes the short translations as shorter sentences tend to have higher BLEU score.  

ROUGE-L calculates a weighted harmonic mean of recall and precision, called F-Score \cite{lin2004rouge}. 
METEOR, on the other hand, is a recall-based metric that measures the effectiveness of the model in grasping the content of reference sentences \cite{banerjee2005meteor}. 

\begin{table*}[h!]
\caption{Comparison of proposed model and simple model with two different architecture. The highest scores are shown in bold.}
\label{table:RQ3}
\small	
\begin{tabularx}{\textwidth}{ p{0.28\textwidth}p{0.09\textwidth}p{0.09\textwidth}p{0.09\textwidth}p{0.09\textwidth}p{0.09\textwidth}p{0.11\textwidth}  }
 \hline
 Models &\multicolumn{6}{c}{Metrics} \\
 \cline{2-7}
 & BLEU1 (\%) & BLEU2 (\%) & BLEU3 (\%) & BLEU4 (\%) & METEOR(\%) &  ROUGE-L (\%) \\
 \hline

{APIContext2Com\textsubscript{GRU}}\textsuperscript{-APIcontext}   &   21.23  & 9.85   &5.45&3.72&9.21&22.83\\

APIContext2Com\textsubscript{GRU}&   \textbf{24.70}  & \textbf{14.45} & \textbf{8.92} & \textbf{5.88} & \textbf{12.13} & \textbf{29.36}  \\

{APIContext2Com\textsubscript{LSTM}}\textsuperscript{-APIcontext}&   21.87  & 10.07   &5.65&3.85&9.76&23.64\\

APIContext2Com\textsubscript{LSTM} &   24.23  & 14.18 & 8.69 & 5.71 & 12.04 & 28.72  \\

 \hline
\end{tabularx}

 \end{table*}

\subsection{RQ1:  The Performance of APIContext2Com Compared to Other Models}

In this research question, we compared our proposed approach and other models. The results are presented in Table \ref{table:RQ1}.  
APIContext2Com outperforms all other models and improve the closest baseline, Rencos, by 1.88 (8.24\%), 2.16 (17.58\%), 1.38 (18.3\%), 0.73 (14.17\%), 1.58 (14.98 \%) and 1.9 (6.92 \%) for BLEU1, BLEU2, BLEU3, BLEU4, METEOR, ROUGE-L, respectively. 
This improvement is due to extracting information from the API context, which is captured by the selection of related APIs. 
Further, adopting proper architecture to separately process each API  has a significant impact on capturing the information. 

Rencos achieves the highest score among baselines, which is reasonable as they incorporate external knowledge by retrieving similar methods. External knowledge as well as exploiting AST structural information causes Rencos to outperform other models. The next results belong to TL-CodeSum and AST-AttendGRU. Similarly, TL-CodeSum incorporates APIs to improve summary generation. However, the result is not as high as expected and it is almost similar to AST-AttendGRU, which  includes code and AST. The reason is that they do not employ any additional information but only API names. These names already exist within the source code, and the information is already processed during source code processing. Therefore, adding only API names might not help to add more knowledge. 
Among all models, DeepCom achieves the lowest scores. We hypothesize that they fail to effectively capture the lexical information of the model as source code tokens are not directly used in the model. 

\subsection{RQ2: The Effect of Ranking Mechanism on the Model's Performance and Best Number of APIs to Include in APIContext2Com}

As some APIs not only are not helpful but are misleading, in this section, we explore additional experiments to discover the optimum number of APIs. As discussed previously, we use a point-based system to rank APIs based on their similarity to the main method. Therefore, the ranking system enables us to select and include the first $n$ top-ranked APIs and discard the rest. 
But, we should decide on the number (i.e., $n$) to include those APIs.
We perform experiments with incorporating different numbers of APIs and eventually discover the optimum API frequency, leading to the highest average scores. We attempted 2, 3, and 4 APIs and as shown in Table \ref{table:RQ2} the model with 3 APIs achieves the highest scores. 
Selecting fewer than three APIs drops the performance by $\approx 0.7$ BLEU-1. Similarly, including more than three APIs decreases the performance, with 4 APIs we achieve lower performance by $\approx 0.3$ BLEU-1. 
The last row in Table \ref{table:RQ2} shows the scores when the ranking system is not used but all APIs are considered. In this case, the performance decreases by $0.8$ score compared to 3 APIs. We conclude that filtering APIs based on the proposed ranking system helps the model to generate comments with higher performance.

Interestingly, though the best number to include in the model is the top-3 APIs, including all APIs or other number of APIs has on par results with 3 APIs. This confirms that other than the ranking system, including the API contexts and the proposed architecture help the model capture relevant information. This is explored in detail in the next RQ.

\subsection{RQ3:  Effect of Incorporating API Context and Distinct RNN Variants (GRU \& LSTM)} \label{sec:gru-effect}

In this RQ, we explore the simple version of the proposed model to better understand the effect of integrating API context. 
The simple model refers to the proposed model excluding API context, receiving only source code and flattened AST as input, shown as APIContext2Com\textsuperscript{-APIcontext}. 
The first row in Table \ref{table:RQ3} represents the performance of APIContext2Com\textsuperscript{-APIcontext} with GRU architecture. APIContext2Com with the API context improves the results of the simple model by $\approx$ 3.47, 4.6, 3.47, 2.16, 2.92, 6.53 for BLEU 1-4, METEOR and ROUGE-L respectively. Therefore, the effect of integrating external knowledge of APIs is not negligible, and it can provide the model with valuable information to predict better comments.


We further explore the effect of different frameworks, RNN variants, by replicating the experiments with LSTM. We replace GRU layers with LSTM and keep other components of our model the same. 
The architecture used in shown as subscript of the model name in Table \ref{table:RQ3}. 
Interestingly, the simple LSTM model, {APIContext2Com\textsubscript{LSTM}}\textsuperscript{-APIcontext}, achieves slightly higher performance compared to {APIContext2Com\textsubscript{GRU}}\textsuperscript{-APIcontext}. But incorporating API improves the GRU variant results more than the LSTM version of the model. Considering BLEU-1 as a representative metric, {APIContext2Com\textsubscript{GRU}}\textsuperscript{-APIcontext} and {APIContext2Com\textsubscript{LSTM}}\textsuperscript{-APIcontext} score are 21.23 and 21.87, respectively. Adding API context,  APIContext2Com\textsubscript{GRU} and APIContext2Com\textsubscript{LSTM} reach 24.7, 24.23, respectively; meaning 3.47, 2.36 improvement over the simple GRU and LSTM models.
This might be related to the GRU framework capable of capturing the API information more effectively as there is more improvement with this architecture when the API context is added.

\subsection{Human evaluation} 
In this study, three metrics, namely, BLEU, ROUGE-L and METEOR, are leveraged to measure the similarity of the reference comments and generated comments by models. However, they sometimes fail to calculate the similarity correctly as they only consider the textual rather than semantic similarity. We, therefore, conduct a human evaluation to acknowledge the quantitative result achieved by the metrics. 

 To perform human evaluation, we use Amazon Mechanical Turk (AMT) website\footnote{https://www.mturk.com/}. It is a crowdsourcing application where processes or tasks can be distributed among distinct remote workers. Tasks are defined as the form of HITs (Human Intelligence Tasks), and remote individuals are paid to perform them. We define a task to compare the similarity of generated comments by four models with the reference comments. The four models are the proposed model and the three best baselines, including Rencos, TL-Codesum, and AST-attendGRU. 
 Similar to previous studies \cite{shahbazi2021api2com}, we select 100 samples randomly from the testing set where each sample contains the reference comment and three generated comments. The evaluators (remote workers) need to give a score to each of generated comments on a 5-point Likert scale, where 1 represents the least semantic similarity between the generated and reference comment, and 5 shows the highest similarity. To attain more accurate and consistent results, each task is rated by three different evaluators and the average value is calculated as the final value. Thus, we would have 300 tasks in total to be done. 
 
The evaluators are educated first by providing examples. The name of the models generating the comments was hidden from the evaluators to provide a fair comparison.
 The result provided by the evaluator is 2.53, 2.39, 2.11, 1.81 for APIContext2Com, Rencos, AST-attendGRU, TL-Codesum respectively. The improvement achieved by APIContext2Com is around 6\%, 20\% and 39\% compared to Rencos, AST-attendGRU, and TL-Codesum, respectively. This result confirms that the generated comments by APIContext2Com are more similar to the ground truth comments than other approaches.

\section{Discussion} \label{sec:discussions}

\subsection{Score Distribution and Effect of API Context}

\begin{figure}[htbp]
\centerline{\includegraphics[width=1\linewidth]{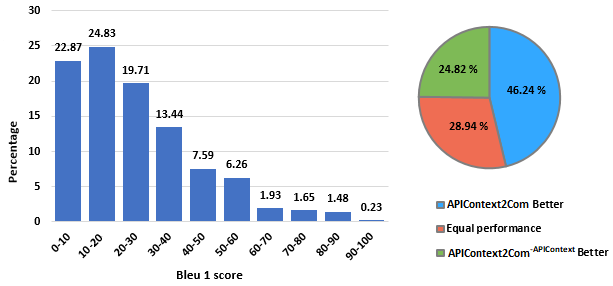}}
\caption{ Effect of API documentation }
\label{fig:discussion-apiEffect}
\end{figure}

In this section, we analyze the performance of the model at the \textit{record} level and also study the effect of integrating the API documentation into the model. For simplicity, only BLEU-1 is considered in the calculations for this part. 

Fig. \ref{fig:discussion-apiEffect}, the bar chart, depicts the breakdown of BLEU-1 scores for APIContext2Com. There is a considerable portion of data, $\sim 47\%$ where score is less than 20. 
This is not surprising as there are a large number of samples that the inputs are not capable of providing any useful information to the model for comment prediction. This could be due to three main reasons. First, the code is not written in a readable way, such as poor naming. Second, the code is fine, and even the model produces a reasonable summary, but the reference comments might not always represent the functionality of code, such as technical debt comments \cite{sharma2022self}. Third, the code and comments are fine, but the model has not seen enough similar examples during training. Approximately 20 percent of samples scores scattered between 20  and 30, which is the average score. Over 32 \% of the samples achieve a score higher than 30, interestingly, sometimes near 100. When investigating these cases, we found many cases with high score where there is an API with a comment which is very similar to the reference comment. 

In Fig. \ref{fig:discussion-apiEffect}, the pie chart explores the effect of adding API documentation to the simple model, APIContext2Com\textsuperscript{-APIcontext} (i.e., only code and AST). The pie chart compares the scores for APIContext2Com\textsuperscript{-APIcontext} and APIContext2Com models at the \textit{record} level. Two models are generating similar comments for $\sim 29\%$ of the samples. 
It is reasonable as these samples mostly do not use any APIs. Over 46\% of cases, the highest portion, receive comments with a better score by APIContext2Com. These are the samples where the API documentation is mostly helpful and relevant.  The remaining samples are the ones that usually carry APIs where the documentation is somehow misleading and the APIContext2Com\textsuperscript{-APIcontext} has higher scores.

\subsection{Code and Comment Length}

\begin{figure}[htbp]
\centerline{\includegraphics[width=1.08\linewidth]{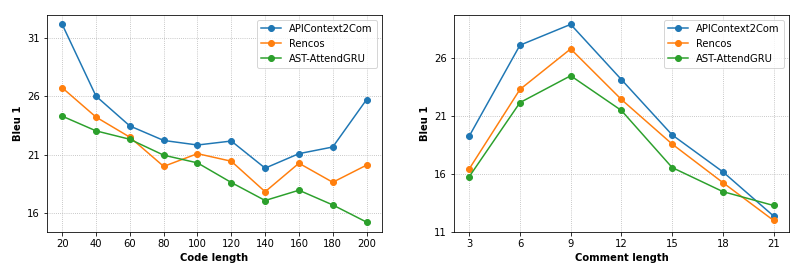}}
\caption{ BLEU score for different lengths of code and comment }
\label{fig:length-analysis}
\end{figure}

A common analysis of the comment generation models is the changes of the BLEU score as the code length or comment length increases. 
Fig. \ref{fig:length-analysis} demonstrates the BLEU-1 score of APIContext2Com and two best baselines, Recos and AST-AttendGRU. The baselines are shown to compare the trends followed by different models. 
The plots show the average scores of code lengths with step size 20 and comment lengths with step size 3.
For example, in the right plot, the score shown for comment length of 6 reflects the average score of the comments with lengths between 3 and 6. 
On average, APIContext2Com has a higher score compared with other models. 
The charts depict a solid improvement for APIContext2Com over different lengths. However, The score of all models declines as the length of code or comments increases. For code length, among all models, APIContext2Com has a steady decrease with a slight positive slope after the length of 140. In contrast, Rencos has a fluctuating trend and AST-AttendGRU performance decrease as the code lengths increase. 
For the comment length, all the models follow a similar trend. The performance increases until the length is 9, and declines after this length. APIContext2Com has more scores for almost all comment lengths compared to Rencos and AST-AttendGRU.

\subsection{Case Study}

\begin{figure}[htbp]
\centerline{\includegraphics[width=1\linewidth]{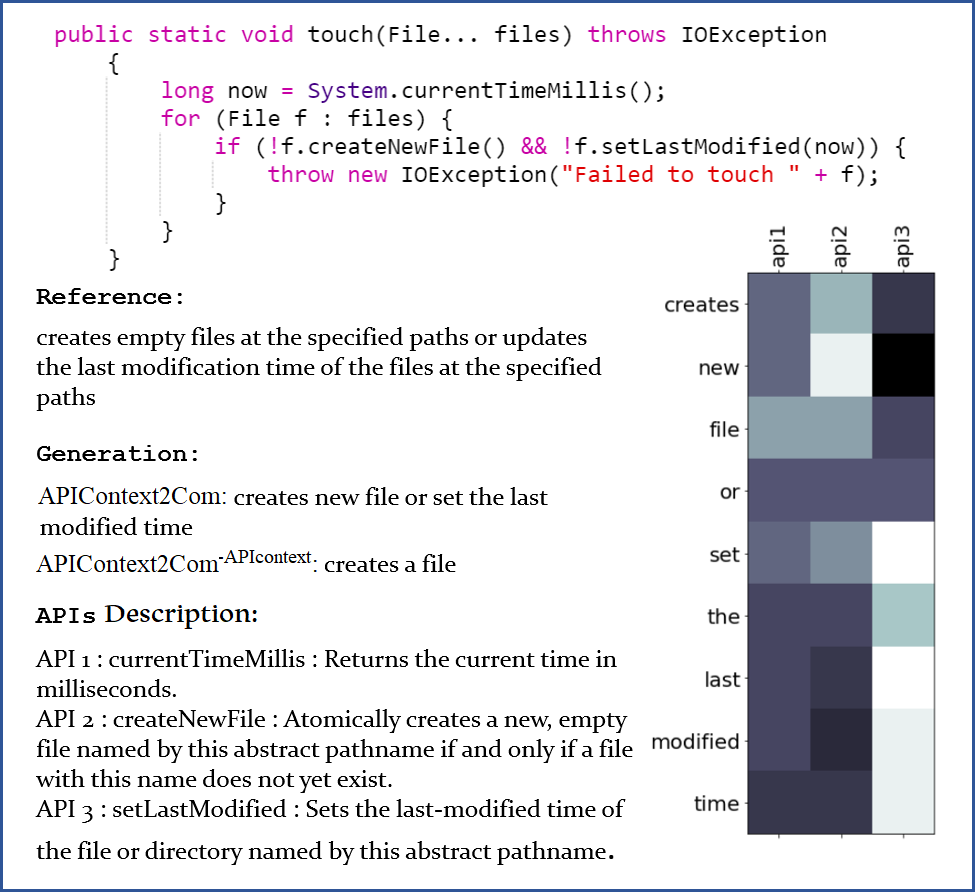}}
\caption{An example where the generated comment is improved when API description is used.}
\label{fig:caseStudyDescription}
\end{figure}

\begin{figure}[htbp]
\centerline{\includegraphics[width=1\linewidth]{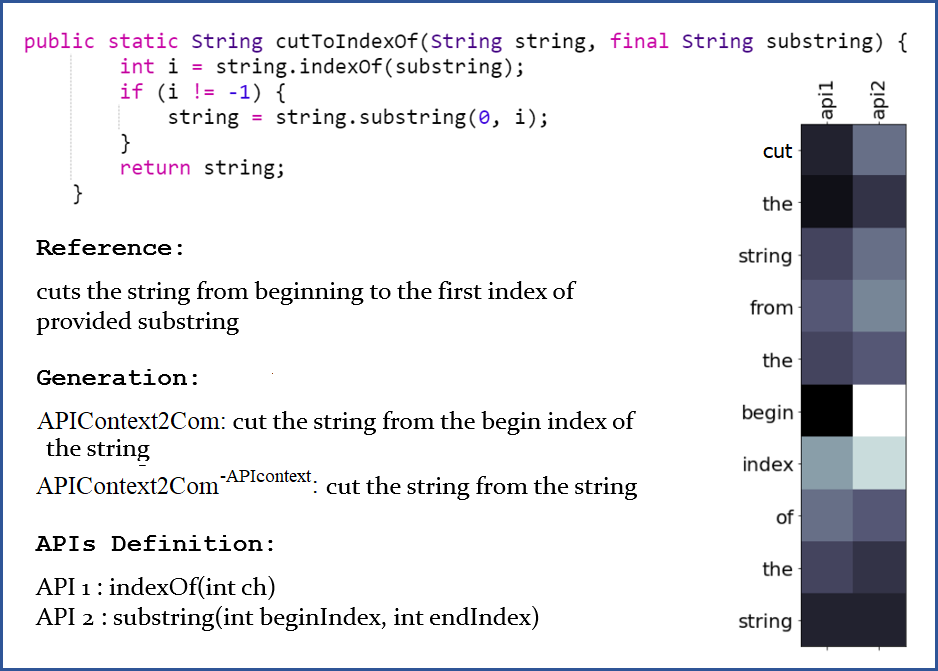}}
\caption{An example of improvement of generated comment when API definition is used.}
\label{fig:caseStudyDefinition}
\end{figure}

In this section, two examples are presented to display how API context enhances the quality of the predicted comment. The first example demonstrates the effect of API description and the second one shows API definition's strength.
In Fig. \ref{fig:caseStudyDescription}, example 1, the comment generated by the simple model, APIContext2Com\textsuperscript{-APIcontext} ``Creates a file" explain limited information in contrast with the reference comment. However, APIContext2Com predicts a more accurate comment, judging from the reference comment. A quick look at the APIs used within this method and their respective comments reveals that API2 and API3, \verb|createNewFile| and \verb|setLastModified|, have very similar descriptions to the reference comment. The heatmap drawn in the figure represents the attention to the APIs description by APIContext2Com.
The lighter color shows more attention. As shown, APIContext2Com attends more to the second API while predicting words \textit{``create new file''} and also attends to the third API while predicting words \textit{``set the last modified time''}. This confirms our hypothesis that API description, when chosen correctly, can help generating comments.


In Fig. \ref{fig:caseStudyDefinition}, a second example is shown, where the generated comments by APIContext2Com\textsuperscript{-APIcontext} model fails to include a critical word ``beginning" in the comment, while the APIContext2com adds this word in the generated comment. A quick look at the code snippet reveals that there is no word similar to ``beginning". Therefore, it is expected from APIContext2Com\textsuperscript{-APIcontext} to miss this word. However, exploring the APIs inside this method shows that a parameter name in API 2's definition is \verb|beginIndex| and the heatmap confirms that the model is attending API2 while predicting ``begin''. 

\section{Threats to validity} \label{sec:threats}


A consideration is designing a neural network model and also reproducing the outputs of the baselines. To confirm the correctness of our code, we double-checked every part. For baselines, we used the public source code provided by the authors and applied the configuration recommended by the authors. 

We adopt a large-scale and well-known Java dataset to ensure the validity of the outputs, but, the finding may not be generalizable to other languages.
More work need to be conducted to understand the effects of API context for other programming languages.

In all the quantitative experiments, the scores are computed based on the metrics in comment generation studies. To mitigate the possible issues introduced by one metric, we included multiple metrics. Besides, we conducted a human evaluation experiment to verify the correctness of the result provided by automatic metrics.

\section{Related works} \label{sec:relatedWork}

Previous studies employed template-based and information retrieval approaches \cite{moreno2013automatic, sridhara2010towards} for the task of automatic code comment generation. In recent years, deep neural networks are applied for code comment generation, Iyer et al. \cite{iyer2016summarizing} being the first one. They proposed an end-to-end encoder-decoder framework, CODE-NN, to generate comments for C\# and SQL. 

Other studies improve the comments by utilizing AST in different ways. 
Hu et al. \cite{hu2020deep} incorporated AST into a seq-to-seq LSTM model to better capture the syntactical information of the source code. 
LeClair et al. \cite{leclair2019neural} combined source code and AST to the model as two different inputs.
Alon et al. \cite{DBLP:conf/iclr/AlonBLY19} proposed an innovative solution to incorporate AST into the model. They select various traversal paths in the AST representation tree and adopt them as different inputs to a LSTM model. 
Adopt AST in tree-based encoders in different frameworks, including Graph neural networks (GNN) \cite{leclair2020improved}, Tree-LSTM \cite{shido2019automatic} and Tree-transformer \cite{DBLP:journals/corr/abs-1908-00449} is used in some works. Fernandes et al. \cite{fernandes2018structured} designed a GNN model, adopting three graphs to represent source code, AST, and lexical use. LeClair et al. \cite{leclair2020improved} introduced a GNN-based model that effectively adapts the AST tree structure (default structure). 
Wan et al. \cite{wan2018improving} incorporated sequential source code text as well as AST into a reinforcement learning model to generate readable comments.
Zhou et al. \cite{zhou2022automatic} proposed a graph attention model to process AST representation and extract the structural information of the code. 

Other approaches consider different aspects to include in the model. 
Nie et al. \cite{nie2022impact} used the timestamps of the source code and summaries and propose a time-segmented evaluation technique. 
Wei et al. \cite{wei2019code} investigated the relation of comment generation and code generation (two opposite tasks) to enhance the performance of the two tasks by developing a dual training framework to integrate duality into attention weights. 
Allamanis et al. \cite{allamanis2016convolutional} introduced a model with convolutional architecture which is able to grasp the long-range attention information. 
Sharma et al. \cite{10.1145/3524610.3527921} proposed a character-based language model to extract the semantic knowledge and a named entity recognition model to grasp the structural information of the model.

Some studies also focus on the evaluation or different aspects of AST. Z{\"u}gner et al. \cite{zugner2021language} employed language-agnostic features such as code snippets and features that can be extracted directly from the AST tree. Tang et. al \cite{tang2022ast} argued AST length is too long to be used directly in the model and it can decrease the model performance. 
Shi et al. \cite{shi2022evaluation} performed a systematic analysis and revealed that different variants of BLEU score can have a significant effect on the model's performance.

Incorporating external knowledge to enhance the performance is another approach found in the literature. 
Wei \cite{wei2019retrieve} and Zhang et al. \cite{zhang2020retrieval} employ information retrieval techniques to retrieve similar code snippets to leverage the corresponding comments. 
EditSum \cite{li2021editsum} proposes a retrieve and edit solution to leverage similar code patterns. 
Wang et al. \cite{wang2020cocogum} included Unified Modeling Language and enclosing class names. 
Haque et al. \cite{haque2020improved} adopted other methods within the same file (file context) where the main method exists as extra inputs. 
Bansal et al. \cite{bansal2021project} extended the level of using external knowledge by combining project context and included other files in the entire project as an additional input. 
Hu et al. \cite{hu2018summarizing} leveraged the name of API sequence to train a API sequence summarization model and then transfer the knowledge to a code comment generation task. 
Shahbazi et al. \cite{shahbazi2021api2com} integrated API documentation into a transformer model. 


Other than RNN-variants, LSTM and GRU, Transformer \cite{vaswani2017attention} is used as the main architecture \cite{shahbazi2021api2com, ahmad2020transformer} or studied \cite{sharma2021thesis} in some recent works.
And finally, another approach is using pre-trained language models (PLM) \cite{feng-etal-2020-codebert, jiang2021treebert, zeng2022extensive, wang2022bridging, 10.1145/3510003.3510050, 10.1145/3524610.3527921, 10.1145/3524610.3527886, hadi_2021_ptm} such as CodeBERT \cite{feng-etal-2020-codebert} where the model is initially pre-trained on a large-scale general-purpose corpus, and is fined-tuned on the target task including comment generation. 

Among all, our work is most similar to Hu's et. al and Shahbazi's et al. \cite{hu2018summarizing, shahbazi2021api2com}. The former uses only API names and the latter incorporates API documentations. Though API documentation is used in \cite{shahbazi2021api2com}, they report a negative result of using API knowledge, especially with larger number of APIs. Our work differs from this study from multiple aspects. We propose a new architecture to encode each API separately, add API definition, and apply a ranking system.

\section{Conclusion} \label{sec:conclusion}
In this study, we proposed a seq-2-seq, multi-encoder code comment generation architecture, namely APIContext2Com.
We used external information regarding pre-defined APIs, API context, as additional inputs and evaluated our model on Java. The API context consists of API description and definition, collected from JDK documentation. 
We further developed a ranking mechanism to filter out APIs which are not helpful in adding knowledge for predicting comments that explains the functionality of the source code. 
Our findings shows an improvement over the state of the arts model, Rencos, and is aligned with our human evaluation results. Our experiments also revealed the effectiveness of using API context. 
In the future, we intend to improve the ranking mechanism and expand the work by leveraging other sources of external knowledge and apply the technique for other programming languages.

\section{Data availability}
We open source our collected API Docs: \url{https://zenodo.org/record/7689869#.Y__K0nbMJD-}

\bibliographystyle{IEEEtran}
{
\balance
\footnotesize
\bibliography{    references}

}




\end{document}